\documentclass[twocolumn,showpacs,preprintnumbers,prl]{revtex4}

\usepackage{epsfig}
\usepackage{latexsym}

\begin{document}  
\title{Excitability mediated by localized structures}
\author{Dami\`a Gomila}
\thanks{Present address: Department of Physics, University of Strathclyde,
Glasgow G4 0NG, Scotland, UK
}
\email{damia@phys.strath.ac.uk}
\affiliation{Institut Mediterrani d'Estudis Avan\c{c}ats (IMEDEA,CSIC-UIB)\\
Campus Universitat Illes Balears, E-07122 Palma de Mallorca, Spain} 

\author{Manuel A. Mat\'{\i}as}
\homepage{http://www.imedea.uib.es/PhysDept/}
\email{manuel@imedea.uib.es}
\affiliation{Institut Mediterrani d'Estudis Avan\c{c}ats (IMEDEA,CSIC-UIB)\\
Campus Universitat Illes Balears, E-07122 Palma de Mallorca, Spain} 

\author{Pere Colet}
\email{pere@imedea.uib.es}
\affiliation{Institut Mediterrani d'Estudis Avan\c{c}ats (IMEDEA,CSIC-UIB)\\
Campus Universitat Illes Balears, E-07122 Palma de Mallorca, Spain} 

\begin{abstract}  
We find and characterize an excitability regime mediated by localized
structures in a dissipative nonlinear optical cavity. The scenario is that
stable localized structures exhibit a Hopf bifurcation to self-pulsating
behavior, that is followed by the destruction of the oscillation in a
saddle-loop bifurcation. Beyond this point there is a regime of excitable
localized structures under the application of suitable perturbations.
Excitability emerges from the spatial dependence since the  system does not
exhibit any excitable behavior locally. We show that the whole scenario is
organized by a Takens-Bogdanov codimension-$2$  bifurcation point.
\end{abstract}

\pacs{42.65.Sf, 05.45.-a, 89.75.Fb}

\date{\today}

\maketitle 

Localized structures (LS) in dissipative media have been found in a variety of
systems, such as chemical reactions, gas discharges or fluids \cite{LS}.
They are also found in optical cavities, due
to  the interplay between diffraction, nonlinearity, driving, and dissipation
\cite{LSoptical,stephan}. These structures, also known
as cavity solitons have to be distinguished from conservative
solitons found for example in propagation in fibers, for which 
there is a continuous family of solutions
depending  on their energy. Instead, dissipative LS are
unique once the parameters of the system have
been fixed. This fact makes this structures potentially useful in optical
(i.e., fast and spatially dense) storage and processing of information
\cite{stephan,FirthScroggieOPN,Coullet1}. 

Localized structures may develop a number of instabilities like start moving,
breathing or oscillating. In the latter case, they would oscillate in time
while remaining stationary in space, like the oscillons found in a vibrated
layer of sand \cite{oscillon}. The occurrence of these oscillons in
autonomous systems has been reported both in optical
\cite{FirthPhSc,josab,longhi} and chemical systems \cite{oscillonve}. 

In the present work we report on a route in which autonomous oscillating 
localized structures are destroyed, leading to a excitability regime.
Excitability is a concept arising originally from biology (e.g. neuroscience),
and it has been found in a variety of systems \cite{Excitreview}, including
optical systems \cite{opticalexcitab,Plaza}.
Typically a system is
said to be excitable if while it sits at an stable fixed point, perturbations
beyond a certain threshold induce a large response before coming back to the
rest state. In addition, excitability is also characterized by the existence of
a  refractory time during which no further excitation is possible. In phase 
space \cite{ErmenRinzel,Izhikevich} excitability occurs for parameter  regimes
where a stable fixed point is close to a bifurcation in which an  oscillation
is created. Possibly the best known example of an excitable system is
FitzHugh-Nagumo model, close to the Hopf bifurcation, although one may also
find excitable behavior mediated by a saddle point, namely either in the form
of an Andronov
(or saddle-node on the invariant circle) 
bifurcation or a saddle-loop (or homoclinic) bifurcation. These three
possible  scenarios are the simplest possible, occurring in systems that,
minimally, can be characterized by two phase variables. 

The concept of excitability has been extended to systems with spatial
dependence by coupling several or many zero-dimensional excitable systems
\cite{Excitreview}. Here we consider a different situation: a system that without
spatial dependence does not show excitable behavior while the localized
structures that appear in the spatially dependent systems do.

\begin{figure}
\centerline{\epsfig{figure=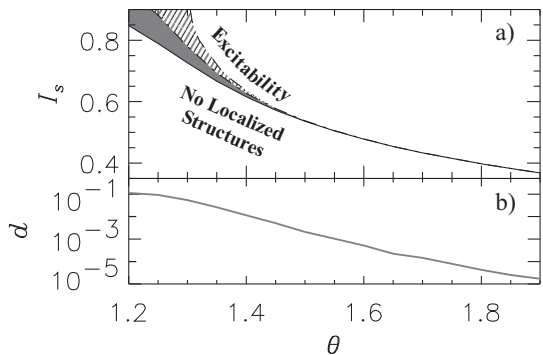,width=0.4\textwidth}}  
\caption{a) Phase diagram: $I_s$ vs. $\theta$.
LS are stable in the shaded region and oscillate in the dashed one. 
Lines indicate bifurcations: saddle-node (solid), 
Hopf (dot-dashed), and sadle-loop  (dashed). 
b) Distance between the saddle-node and Hopf lines.} 
\label{phased} 
\end{figure}

A prototype model in nonlinear optics displaying the formation of LS is the
one introduced by Lugiato and Lefever for an optical cavity filled with a Kerr medium
\cite{Lugiato-Lefever}. In the mean field approximation the dynamics
of the scaled slowly varying amplitude of the complex  field $E(\vec{x},t)$ is
given by
\begin{eqnarray} 
\frac{\partial E}{\partial t} = - (1+ i\theta) E + i\nabla^2
E + E_0 + i\left| E \right|^2 E , 
\label{LLeq} 
\end{eqnarray} 
where $\nabla^2 = \partial^2/\partial x^2 + \partial^2/\partial y^2$. The first
term describes cavity losses, $E_0$ is a homogeneous (plane wave) driving
field, and $\theta$  the cavity detuning with respect to $E_0$. 
Eq.~(\ref{LLeq}) has been normalized by the cavity decay rate, and it
has a homogeneous steady state solution $E_s$ given implicitly by
$E_s=E_0/(1+i(\theta-I_s))$, where $I_s=|E_s|^2$. We use the intra-cavity
background  intensity $I_s$ together with $\theta$ as  convenient control
parameters. The homogeneous solution has a instability leading to the formation
of a hexagonal pattern at $I_s =1$.  The bifurcation starts subcritically and
the pattern coexists with the homogeneous solution
\cite{Lugiato-Lefever,Scroggie}, so that a  stable-unstable pair of LS is
created in a saddle-node  bifurcation \cite{Coullet1}. The unstable branch (the
so-called lower or middle branch) has a single unstable direction.
The region of existence of these LS, also known as Kerr cavity solitons, has
been characterized in  \cite{FirthPhSc,josab}, and is partially shown in Fig.
\ref{phased}. Increasing $\theta$, the LS undergoes a supercritical
Hopf bifurcation and starts to oscillate autonomously 
\cite{FirthPhSc,josab,Skryabin}. For one spatial dimension, Eq. (\ref{LLeq}) also 
exhibits LS in the appropriate parameter regime, but these structures
do not undergo any Hopf bifurcation.

\begin{figure}
\centerline{\epsfig{figure=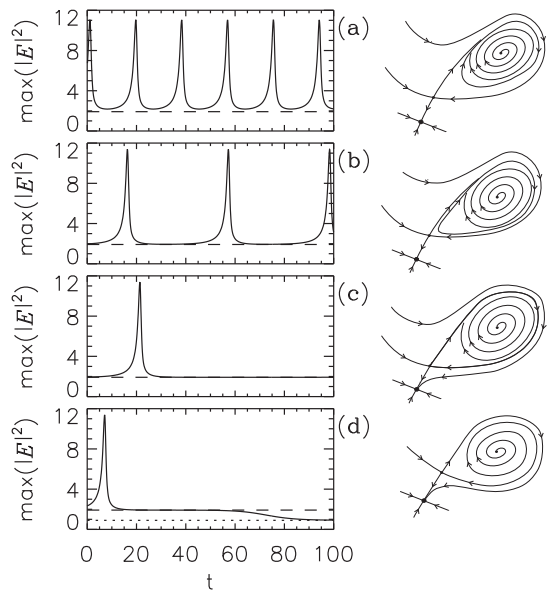,width=0.4\textwidth}}
\caption{Left: LS maximum intensity as a function of time
for increasing values of the detuning parameter $\theta$. From top to
bottom $\theta=1.3, 1.3047, 1.30478592, 1.304788$. $I_s=0.9$. Right: Sketch of
the phase space for each parameter value. The thick line shows
the trajectory of the LS in phase space.}
\label{fig:homoclinic}
\end{figure}

As the control parameter $\theta$ is further increased part of the limit cycle
moves closer and closer to the lower-branch LS, which is a saddle point in 
phase space, as illustrated in Fig.~\ref{fig:homoclinic}. On the left column we
plot the time evolution of the LS maximum obtained from numerical integration
of Eq.~(\ref{LLeq}), the dashed line shows for comparison the maximum of the
lower-branch LS. On the  right column we sketch the evolution on phase space
projected on two variables.  At a certain critical value a global bifurcation
takes place: the cycle touches the lower-branch LS and becomes a homoclinic
orbit (Fig.~\ref{fig:homoclinic}c). 
This is an infinite-period bifurcation called {\it saddle-loop} or {\it
homoclinic  bifurcation} \cite{Gaspard}. For
$\theta > \theta_c$, the saddle connection breaks and the loop is  destroyed
(Fig.~\ref{fig:homoclinic}d). After following a trajectory in phase space close
to the previous loop the LS approaches the saddle point (dashed line) where
the evolution is dominated by its slow stable manifold 
(see the long plateau between $t=15$ and
$t=60$ in Fig.~\ref{fig:homoclinic}d). Finally  the LS decays to the
homogeneous  solution (dotted line). For larger values of $\theta$ the
trajectory moves away from the saddle and, therefore, the decay to the
homogeneous solutions takes place in shorter times.

\begin{figure}
\centerline{\epsfig{figure=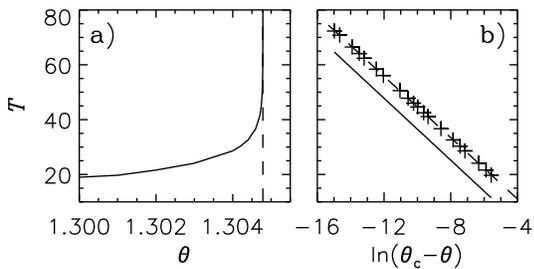,width=0.4\textwidth}}
\caption{a) Period of the limit cycle $T$ as a function 
of the detuning $\theta$ for $I_s=0.9$. The vertical dashed line indicate the 
threshold of the saddle-loop bifurcation $\theta_c=1.30478592$. b) Period
$T$ as a function of $\ln(\theta_c-\theta)$. Crosses correspond to numerical
simulations while the solid line has a slope $1/\lambda_1$ with 
$\lambda_1=0.177$ obtained from the linear stability analysis of the
lower-branch LS.}
\label{fig:scaling}
\end{figure}

The saddle-loop bifurcation has a characteristic {\it scaling law} that governs
the period $T$ of the limit cycle as the bifurcation is approached. Close to the
critical point the system spends most of the time close to the unstable LS
(saddle). $T$ can be then estimated by the
linearized dynamics around the saddle \cite{Gaspard} 
\begin{equation}
T\propto -{1 \over \lambda_1}\ln(\theta_c-\theta),
\label{powerlaw}
\end{equation} 
where $\lambda_1$ is the unstable eigenvalue of the saddle point. We are now
going to show that this scaling law is verified in our system.
Fig.~\ref{fig:scaling} shows the period of the LS limit cycle as a function of
the control parameter $\theta$. As expected, the period of the limit cycle
diverges logarithmically as the bifurcation is approached. We then evaluate
$\lambda_1$ with arbitrary precision from a semi-analytical stability analysis 
of the unstable LS as in \cite{josab}. The  lower-branch
LS has one single positive eigenvalue $\lambda_1=0.17713581$. In
Fig.~\ref{fig:scaling} (right) we plot using crosses the period of the
oscillation LS as a function of $ln(\theta_c-\theta)$ obtained from numerical
simulations of Eq.~(\ref{LLeq}). Performing a linear fitting we obtain a slope
$5.60$, in excellent agreement with the theoretical prediction
$1/\lambda_1=5.65$, proving the existence of a
saddle-loop bifurcation for the oscillating LS. We should note that
the theory was developed for planar bifurcations, therefore, as the phase space
is a plane, the saddle has one unstable direction and one 
attracting direction \cite{Gaspard}. The stable manifold of the 
unstable LS is, however,  infinite dimensional. The success of the planar
theory to describe our infinite dimensional system can be attributed to the
fact that, somehow, the dynamics of the LS is effectively two-dimensional
with a single unstable manifold and an effective stable manifold.

In systems without spatial dependence it has been shown that an scenario
composed by a saddle-loop bifurcation and a stable fixed point leads to a
excitability regime \cite{ErmenRinzel,Izhikevich,Plaza}.
In our infinite-dimensional system LS does indeed show en excitable behavior: 
the homogeneous solution is a globally attracting fixed point
but localized disturbances (above the lower-branch LS)
can send the system on a long excursion through  phase space before returning
to the fixed point. 

\begin{figure}
\centerline{\epsfig{figure=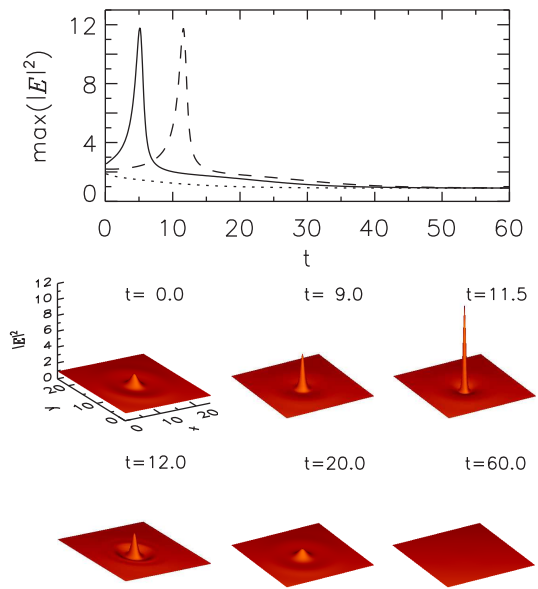,width=0.4\textwidth}}
\caption{Evolution starting from the homogeneous solution
($I_s=0.9$) plus a localized perturbation of the form of the
unstable LS multiplied by a factor $a$. Top panel shows the time evolution
of the maximum intensity for $a=0.8$ (dotted line); $a=1.01$ (dashed line); 
$a=1.2$ (solid line). The $3D$ plots show the transverse profile at different
times for $a=1.01$.
}
\label{exctraj} 
\end{figure}

 Fig.~\ref{exctraj} shows the  resulting trajectories of applying a perturbation
in the direction of the unstable LS with three different intensities: one
below the excitability threshold (dotted line), and two above: one very close
to threshold (dashed line) and the other well above (solid line). In the first
case the system relax exponentially to the homogeneous solution, while in the
latter two cases it perform a long  excursion before returning to the stable
fixed point. In the case of a near threshold excitation the refractory period
is appreciably longer due to the effect of the saddle. 
The spatial profile of the localized structure is shown in Fig.~\ref{exctraj}.
The peak grows to a large value until the losses stop it. Then it decays
exponentially  until it disappears. A remnant wave is emitted out of the center
dissipating the remaining energy.

In absence of spatial degrees of freedom, Eq.~(\ref{LLeq}) does not present any
kind of excitability. This behavior is strictly related to the dynamics of the
2D LS. Eq.~(\ref{LLeq}) is, in fact, a nonlinear Schr\"odinger equation (NLSE)
with driving and damping. The  phenomenon of self-focusing collapse
\cite{NewellAkhmed} of the 2D NLSE is behind the long  excursion in
phase space. When a localized perturbation concentrates enough power, the
self-focusing nonlinear mechanism induces a concentration of energy at that
place. In the absence of losses a collapse, i.e. a divergence of the energy
concentration, would take place in a  finite time \cite{collapse}. However, the
presence of losses prevents this  collapse \cite{losses}. The perturbation is
finally dissipated and the system returns to the homogeneous solution. The same
mechanism has also been used to explained the transport properties and
dissipation rates in a wide class of  turbulent flows \cite{Newell}. The
mechanism leading to excitability proposed  in this letter is therefore not
restricted to nonlinear optics. It may have implications in plasma physics and
hydrodynamics, where coherent structures may have similar features as LS.
Furthermore, LS appearing on different systems may undergo
a Hopf bifurcation due to the nonlinear dynamics, even in the 1D case, 
as for example in 
parametrically amplified optical systems \cite{longhi}.

In the limit of large detuning, the saddle-node, Hopf and saddle-loop 
bifurcation lines meet asymptotically, at $I_s=0$ as shown in
Fig.~\ref{phased}. It is known that the intersection of a saddle-node line with
a Hopf line is a Takens-Bogdanov (TB) codimension-2 bifurcation point
\cite{Guckenheimer} iff there is a double zero eigenvalue (the imaginary part 
of the Hopf vanishes as approaching the intersection with the saddle-node). The
unfolding around a TB point leads to a saddle-loop bifurcation line
\cite{Guckenheimer}. So, this unfolding fully explains the observed scenario,
where once again our formally infinite-dimensional system appears to be
perfectly described by a dynamical system in the plane. 

\begin{figure}
\centerline{\epsfig{figure=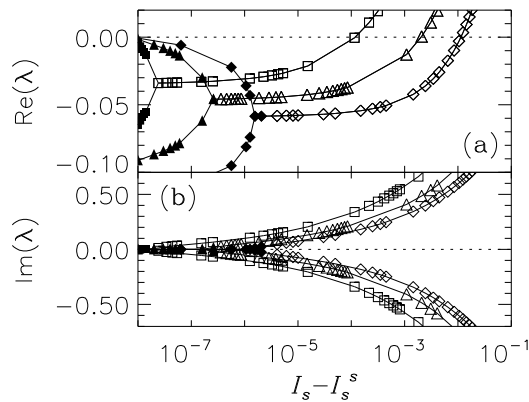,width=0.4\textwidth}}
\caption{Real (a) and imaginary (b) parts of the
stable LS eigenvalues for three vertical cuts in Fig.
\ref{phased} corresponding to detuning $\theta$:
$1.7$ ($\Box$); $1.5$ ($\triangle$); $1.4$ ($\diamondsuit$) versus the difference between
$I_s$ and its value at the saddle-node bifurcation $I_s^s(\theta)$.}
\label{fig:values}
\end{figure}

In Fig.~\ref{fig:values} we plot the two eigenvalues with largest real part of the
LS for parameter values corresponding to  three vertical cuts of
Fig.~\ref{phased}a). Open symbols indicate eigenvalues with a non-zero
imaginary part while filled symbols correspond to real eigenvalues.
Where open symbols cross zero in Fig. \ref{fig:values}a) corresponds to
the Hopf bifurcation while filled symbols crossing zero
indicates the saddle-node bifurcation. For the three plots we have taken
as origin the saddle-node bifurcation.  The important point is that in the
line  of the two complex conjugate eigenvalues responsible of the Hopf 
bifurcation there is a place where the imaginary part vanishes leading to two
real eigenvalues, the largest of which is precisely the responsible of the
saddle-node bifurcation. As detuning increases the Hopf and saddle-node
bifurcation points gets closer and closer but the structure of eigenvalues
remains unchanged so that when the Hopf and saddle-node bifurcation will 
finally meet the Hopf bifurcation will have zero frequency, signaling a TB 
point.

The TB point occurs asymptotically in the limit of large detuning $\theta$ 
and small pump $E_0$. This limit corresponds to the 
case in which Eq.~(\ref{LLeq}) becomes the conservative NLSE \cite{FirthLord}. 
Details of the Hopf instability in this limit were studied in \cite{Skryabin}, 
where there is evidence of the double-zero
bifurcation point, but the unfolding leading to the scenario presented here 
is not analyzed. 

In conclusion, we have analyzed an excitable regime associated to the existence
of localized structures. We thus show that, in order to
exhibit excitability, extended systems do not necessarily require to have local
excitable behavior, instead such phenomenon can emerge due to the spatial 
dependence through the dynamics of a coherent (localized) structure. This
open the possibility to observe excitable behavior in a whole new class of
systems where excitability was not thought to be present. Coherent resonance
\cite{Pikovsky-Kurths} is also  expected to be observed in theses systems upon
application of localized  disturbances of stochastic amplitude. Finally,
localized structures have been shown to have a great potential as bits for
optical memories \cite{stephan,FirthScroggieOPN}. This new excitable regimen opens also a new
possibility for the use of  localized structures as centers to process optical
information in a similar  way neurons do with electrical signals.

We thank G. Orriols for useful discussions.
We acknowledge financial support from MEC (Spain) and FEDER:
Grants BFM2001-0341-C02-02, FIS2004-00953, and
FIS2004-05073-C04-03. DG acknowledges financial
support from EPSRC (GR/S28600/01).

\end{document}